\documentclass{pasa}%

\usepackage{graphicx}

\title[Self-similar solution of advection-dominated discs]{Self-similar structure of resistive ADAFs with outflow and large-scale magnetic field}

\author[S.M. Ghoreyshi]{S.M. Ghoreyshi\thanks{smghoreysi64@gmail.com}
\affil{Department of Physics, Faculty of Sciences, Golestan University, Gorgan 49138-15739, Iran.}%
}%

\jid{PASA}
\doi{10.1017/pas.\the\year.xxx}
\jyear{\the\2020}

\usepackage{aas_macros}
\usepackage{hyperref}
\hypersetup{colorlinks,citecolor=blue,linkcolor=blue,urlcolor=blue}

\begin{document}

\begin{frontmatter}
\maketitle

\begin{abstract}
The observations and simulations have revealed that large-scale magnetic field and outflows can exist in the inner regions of an advection-dominated accretion disc where the resistive diffusion may also be important. In the present paper, the roles of large-scale magnetic field and outflows in the structure of resistive advection-dominated accretion discs are explored by assuming that the accretion flow is radially self-similar. In the non-ideal magnetohydrodynamic (MHD) approximation, the results show that the angular velocity is always sub-Keplerian when both the outflow and the large-scale magnetic field are taken into account. A stronger toroidal field component leads to faster rotation, while the disc rotates with faster rate if the vertical field component is weaker. The increase of magnetic diffusivity causes the infall velocity to be close to Keplerian velocity. Although the previous studies in the ideal MHD approximation have shown that the disc temperature decreases due to the vertical field component, we find that the effect of vertical field component on the temperature of a resistive disc depends on the magnetic diffusivity. When the magnetic diffusivity is high, the more efficient mechanism for decreasing the disc temperature can be the outflows, and not the large-scale magnetic field. In such a limit of the magnetic diffusivity, the components of the large-scale magnetic field enhance the gas temperature. The increase of temperature can lead to heating and acceleration of the electrons and help us to explain the origin of phenomena such as the flares in Sgr A*. On the other hand, the infall velocity in such a limit rises as the temperature increases, and therefore the surface density falls to too low values. Any change in the density profile can alter the structure and the emitted spectrum of disc.
\end{abstract}

\end{frontmatter}

\section{INTRODUCTION }
\label{sec:intro}

Accretion flows can be divided into several classes based on the accretion rate. The standard accretion disc model \citep{Shakura1973} which belongs to low regime cannot be applied to systems such as Sgr A* \citep{Kato2008}. To explain the spectral energy distributions of accreting sources with very low and high accretion rates, respectively, the optically thin advection dominated accretion flows (ADAFs) \citep{Ichimaru1977,Narayan1994} and slim discs (or optically thick ADAFs) \citep{Abramowicz1988} were introduced. In these accretion flows, the energy released due to dissipation processes is not radiated away and is retained within the accreting fluid. Instead, this internal energy can be advected radially inward. ADAFs can be appropriate models to sources like the ultraluminous X-ray sources \citep{Godet2012,Mondal2019}, low-luminosity AGNs \citep{Park2019,Younes2019}, narrow-line Seyfert 1 galaxies \citep{Haba2008,Meyer2011}, and Sgr A* \citep{Yuan2002,Yuan2014}.

The numerical simulations have revealed that there exist considerable outflows in the optically thin or thick ADAFs \citep{Ohsuga2009,Yuan2012a,Hashizume2015}. This result has been confirmed by the observations of low-luminosity AGNs, Seyfert 1 galaxies, and X-ray binaries \citep{Crenshaw2012,Homan2016,Park2019}. The outflows by removing mass, angular momentum and energy from the disc \citep{Pudritz1985,Knigge1999,Xue2005,Cao2016,Cao2019} can strongly affect the structure of accreting sources. For example, the gas can be pushed away from the disc surrounding the black hole by the outflows. This extraction causes the accretion rate of the black hole to decrease \citep{Yuan2018,Bu2019ApJ}. \cite{Blandford1999} by using a global analytical solution demonstrated that the radial-dependence of the accretion rate is as a power-law function. Their power law index has been assumed to be in the range from 0 to 1. The numerical simulations have also shown that the accretion rate in the presence of outflows decreases as such a power-law function \citep[e.g.][]{Ohsuga2005,Yuan2012a,Yuan2012,Bu2013,Yang2014,Bu2018}. Although recent theoretical studies showed that the power index can lie between 0.4 to 1.0 \citep{Yuan2012,Bu2013,Yang2014}, the observations of Sgr A* combined with the radiatively inefficient accretion model have predicted a range $0.3<s<0.4$ \citep{Yuan2003}.

Different mechanisms have been suggested that drive the outflows \citep[e.g.][]{Chelouche2005,Ohsuga2011,Cao2014,Yuan2015,Hashizume2015}. One of them can be large-scale magnetic fields \citep{Blandford1982,Igumenshchev2003,Cao2014,Li2019}. The magnetic fields by transferring the angular momentum of a magnetized disc due to the magnetic braking \citep{Stone1994} or magnetorotational instability (MRI) \citep{Balbus1998} can also modify the disc structure. Several attempts to examine the roles of outflows and large-scale magnetic fields in ADAFs have been made over the last several years. When large-scale magnetic fields with only toroidal component are considered, self-similar solutions indicate that this component of magnetic field rises the gas temperature of an ADAF with outflow \citep[e.g.][]{Abbassi2010,Ghasemnezhad2016,Bu2019Univ,Ghoreyshi2020}. Such high temperatures do not match values obtained from fitting the observational data of Seyfert galaxies \Citep[see][]{Yuan2004}. Thus, the effect of other components of the field on the structure of ADAFs must be investigated. The influence of poloidal field component in an ADAF in the absence of outflows has been examined by \cite{Zhang2008}. Adopting the ideal magnetohydrodynamic (MHD) approximation, they concluded that the poloidal field component can reduce the temperature \citep[see also][]{Ghasemnezhad2017}. But \cite{Bu2009} showed that the role of poloidal field component in the temperature reduction can be more significant when the outflows are considered \citep[see also][]{Mosallanezhad2013}.

Although most of the theoretical studies of accretion discs use the ideal MHD approximation, the resistive diffusion of magnetic field plays a key role in systems such as the protoplanetary \citep[e.g.][]{Sano1998,Bai2013,Meheut2015,Flock2017} and the dwarf nova \citep[e.g.][]{Gammie1998,Sano2003,Scepi2018} discs. Numerical simulations have shown that the resistive dissipation can affect the growth rate and the saturation level of MRI \citep{Fleming2000,Sano2002a,Sano2002,Masada2008,Simon2009,Bai2013,Rodgers2016}, and also the rate of angular momentum transport \citep{Ziegler2001,Fromang2007,Longaretti2010} in these systems. The importance of resistivity in other systems like the discs around Kerr black holes \citep{Kudoh1996,Qian2017,Qian2018,Shaghaghian2020}, and even Sgr A* at the Galactic Center \citep[e.g.][]{Melia2001} has also been recognized. In hot accretion flows modeled for Sgr A*, the Coulomb coupling between ions and electrons is not strong enough. Since these flows, such as ADAFs, have the mean free path for Coulomb interactions larger than the typical size of the system, they are effectively collisionless \citep{Quataert1998} and the ideal MHD approximation may not be sufficient to study their properties. The recent studies have shown that the resistive dissipation in hot accretion flows around black holes can modify the amount of magnetic field and the role of magnetic field in transferring the angular momentum \citep{Zeraatgari2018}. \cite{Zeraatgari2018} suggested that the magnetic dissipation heating in regions of disc where outflows may be important is comparable to the viscous heating. Hence, the magnetic dissipation heating for increasing the electron temperature of ADAFs may be an effective agent \citep{Bisnovatyi1997,Ding2010}. On the other hand, the astrophysical outflows depend on the strength of magnetic diffusivity and type of the field component \citep{Kuwabara2000,Fendt2002,Igumenshchev2003,Cemeljic2014,Qian2018,Vourellis2019}. Thus, one can expect that the structure of accretion discs is modified through the presence of resistive diffusion. \cite{Faghei2012MNRAS} showed when the resistive accretion flows with outflow is investigated, the disc has higher temperature due to magnetic diffusivity. Perhaps this result arises from considering a purely toroidal magnetic fields. Although \cite{Abbassi2012ApSS} studied the effect of other field components on the dynamics of resistive ADAFs, the role of outflows was neglected in their work. In the present paper, we shall extend the work of \cite{Faghei2012MNRAS} by considering other field components. Indeed, our main aim is to investigate the roles of outflows and large-scale magnetic fields in restive ADAFs.

Here, the large-scale magnetic fields are assumed to have both $z$ and $\phi$ components. In order to study the effects of outflows and large-scale magnetic fields, we present the self-similar solutions of restive ADAFs. The paper is organized as follows. We first formulate basic equations for such a disc in Section 2 when outflows exist. In Section 3, the basic equations are solved using self-similar method and their results are discussed in Section 4. We then summarize our conclusions in final section.

\section{Basic Equations}

To investigate the structure of resistive ADAFs, we first write the basic equations in the cylindrical coordinate system $(r,\phi, z)$ centered on the central object. The accretion disc is assumed to be stationary ($\partial/\partial t=0$) and axisymmetric ($\partial/\partial\phi=0$). All flow variables are considered to be a function only of $r$. We have ignored the relativistic effects and used Newtonian gravity. Although the dominant component of magnetic field in the main disc body, inner torus, and corona is in the $\phi$-direction, the magnetic field near the poles may be poloidal, especially with the vertical component \citep{Hirose2004}. Therefore, we adopt a large-scale magnetic field with the components $B_\phi$ and $B_z$, and neglect the radial component $B_r$. Hence, the basic equations of our model are written as
\begin{equation}\label{eq:continuity}
\frac{d}{dr} (r \Sigma V_r)+\frac{1}{2\pi} \frac{d \dot{M}_W}{dr}=0,~~~~~~~~~~~~~~~~~~~~~~~~~~~~~
\end{equation}

\begin{eqnarray}\label{eq:radial-momentum}
V_r \frac{d V_r}{dr}=\frac{{V_\phi}^2}{r}-\frac{GM_*}{r^2}-\frac{1}{\Sigma}\frac{d}{dr}(\Sigma c_s^2)~~~~~~~~~~~~~~~~
\nonumber\\~~~-\frac{1}{2\Sigma}\frac{d}{dr}\Big(\Sigma ({c_\phi}^2+{c_z}^2)\Big)-\frac{{c_\phi}^2}{r},
\end{eqnarray}

\begin{equation}\label{eq:azimuthal-momentum}
\frac{V_r}{r} \frac{d}{dr} (r V_{\phi})=\frac{1}{r^2\Sigma}\frac{d}{dr}( r^3\nu\Sigma \frac{d\Omega}{dr})-\frac{\ell^2 \Omega }{2\pi\Sigma} \frac{d\dot{M}_W}{dr},~~~~
\end{equation}

\begin{equation}\label{eq:energy}
\frac{\Sigma V_r}{\gamma-1}\frac{d c_s^2}{dr}-2HV_r c_s^2\frac{d\rho}{dr}=f(Q_{\rm vis}+Q_B)-Q_W,
\end{equation}

\begin{equation}\label{eq:azimuthal-induction}
\dot{B}_\phi=-\frac{d}{dr}(V_r B_\phi)+\eta\frac{d}{dr}\Big(\frac{1}{r}\frac{d}{dr}(rB_\phi)\Big),~~~~~~~~~~~~~~~~~~~~
\end{equation}

\begin{equation}\label{eq:vertical-induction}
\dot{B}_z=-\frac{1}{r}\frac{d}{dr}(rV_r B_z)+\frac{\eta}{r}\frac{d}{dr}(r\frac{dB_z}{dr}),~~~~~~~~~~~~~~~~~~~~
\end{equation}
where the surface density $\Sigma$ is defined as $\Sigma=2\rho H$, and $\rho$ is the midplane density of the disc. Here, the radial infall velocity and the rotational velocity of the disc are denoted as $V_r(<0)$ and $V_\phi(=r\Omega)$, respectively. $\dot{B}_{\phi,z}$ are the field escaping/creating rate which may be due to the magnetic diffusion or dynamo effect \citep{Oda2007}. The sound speed $c_s$, and the Alfv\'{e}n sound speeds $c_{\phi,z}$ are also defined as
\begin{center}
${c_s}^2=p/\rho,~{c_{\phi,z}}^2={B_{\phi,z}}^2/ 4\pi\rho,$
\end{center}
where $p$ is the gas pressure in the disc. The hydrostatic balance in the vertical direction leads to a relation between $H$ and $c_{s,\phi}$. We have
\begin{equation}\label{eq:hydrostatic}
\frac{GM_*}{r^3}H^2={c_s}^2(1+\beta_\phi).~~~~~~~~~~~~~~~~~~~~~
\end{equation}
Here, $M_*$ is the mass of the central object. In this paper, we use definitions $\beta_{\phi,z}=(1/2)(c_{\phi,z}/c_s)^2$. For the hot accretion flows, the typical values of $\beta_\phi$ and $\beta_z$ may be between 0.01 and 1 \citep[e.g.][]{DeVilliers2003,Beckwith2008}. However, the numerical simulations indicate that the accretion flows in the hot accretion disc change from a gas pressure-supported state to a magnetically supported state as soon as the thermal instability arises \citep{Machida2006}. Under these conditions, one would expect that $\beta_\phi$ and $\beta_z$ exceed unity. Hence, we shall also consider the cases with values greater than 1.

The mass-loss rate by outflow $\dot{M}_W$ is
\begin{equation}\label{eq:mass-loss}
\dot{M}_W(r)=\int 4\pi r'\dot{m}_W(r') dr',~~~~~~~~~~~~~~~~~~~
\end{equation}
where $\dot{m}_W$ is mass loss rate per unit area from each disc face. The definition of the accretion rate, i.e., $\dot{M}=-2\pi r\Sigma V_r$, and equation (\ref{eq:continuity}) can result in
\begin{equation}\label{eq:accretion-rate}
\frac{d \dot{M}}{dr}=\frac{d \dot{M}_W}{dr}.~~~~~~~~~~~~~~~~~~~~~~~~~
\end{equation}
If the radial dependence of $\dot{M}$ is $\dot{M}_0(\frac{r}{r_0})^s$ \citep{Blandford1999} where $\dot{M}_0$ is the mass accretion rate at the outer boundary $r_0$, we have
\begin{equation}
\dot{m}_W=\frac{s}{4\pi r_0^2}\dot{M}_0(\frac{r}{r_0})^{s-2}.~~~~~~~~~~~~~~~~~~~~~
\end{equation}
Here, $s$ represents the outflow strength in the disc. Note that $s = 0$ corresponds to a disc without winds.

The last term of right hand side in equation (\ref{eq:azimuthal-momentum}) represents the angular momentum extracted by the wind where $\ell$ is the length of the lever arm. In the present paper, $\ell$ is assumed to be equal to 1 describing the case in which outflowing material carries away the specific angular momentum it had at the point of ejection \citep{Knigge1999}. The cases with $\ell < 1$ (or $>1$) are expected to describe the outflows that carry away less (or more) angular momentum.

In the energy equation (\ref{eq:energy}), the advection factor $f$ lying between 0 and 1 is defined as $f =1 - \big(Q_{\rm rad}/Q^{+}\big)$ in which $Q^{+}=Q_{\rm vis}+Q_B$. Here, $Q_{\rm rad}$, $Q_{\rm vis}$, and $Q_B$ are the radiative cooling rate, the viscous heating rate, and the resistive heating rate, respectively. The viscous and resistive heating rates, respectively, are
\begin{center}
    $Q_{\rm vis}=\nu\Sigma(r\frac{d \Omega}{dr})^2,$
\end{center}
and
\begin{center}
    $Q_B=\frac{H\eta}{2\pi}\mid\nabla\times\mathbf{B}\mid^2.$
\end{center}
Note that $f$ in the advection-dominated regime is equal to unity. In order to obtain the energy loss due to outflow, the wind is assumed to be driven from the disc surface. If some of the energy generated by viscosity (or any other mechanism) in the disc may power the wind, it must supply some or all of the wind's binding and kinetic energies. \cite{Knigge1999} showed that the energy loss by outflow $Q_W$ is defined by
\begin{center}
    $Q_W=\frac{1}{2}\zeta\dot{m}_W {V_K}^2,$
\end{center}
where $V_K$ ($=r\Omega_K=\sqrt{GM_*/r}$) is the Keplerian speed and the efficiency factor $\zeta$ may be a function of $\ell$. The efficiency factor for a case with $\ell=1$ is equal to 1 \citep{Knigge1999}.

The magnetic diffusivity $\eta$ is assumed to be
\begin{equation}\label{eq:eta}
\eta=\frac{\nu}{P_m}=\frac{\alpha c_s H}{P_m}.~~~~~~~~~~~~~~~~~~~~~~~~~
\end{equation}
Here, $P_m$ is the magnetic Prandtl number. Although the magnetic Prandtl number is generally considered to be a constant, its value depends on the disc radius \citep{Balbus2008}. The studies show that the magnetic Prandtl number may lie in the range from $10^{-3}$ to $10^3$ in the discs around compact X-ray sources and AGNs \citep{Balbus2008}. Note that we use the $\alpha$-prescription of \cite{Shakura1973} for the kinematic viscosity coefficient $\nu$.

\section{Self-similar solutions}

In this paper, some of parameters such as the magnetic Prandtl number are assumed to be a constant for simplicity. We employ a self-similar treatment similar to \cite{Shadmehri2008} and \cite{Faghei2012MNRAS}. Our self-similar solutions are
\begin{equation}\label{eq:s}
\Sigma(r)=C_0\Sigma_0(\frac{r}{r_0})^{s-\frac{1}{2}},~~~~~~~~~~~~~~~~~~~~~~
\end{equation}

\begin{equation}\label{eq:v}
V_r(r)=-C_1\sqrt{\frac{GM_*}{r_0}}(\frac{r}{r_0})^{-1/2},~~~~~~~~~~~~~~~~
\end{equation}

\begin{equation}\label{eq:o}
V_{\phi}(r)=C_2\sqrt{\frac{GM_*}{r_0}}(\frac{r}{r_0})^{-1/2},~~~~~~~~~~~~~~~~~
\end{equation}

\begin{equation}\label{eq:cs}
c_s^2(r)=C_3\frac{GM_*}{r_0}(\frac{r}{r_0})^{-1},~~~~~~~~~~~~~~~~~~~~~~~~~~~~~
\end{equation}

\begin{equation}\label{eq:cphi}
c_{\phi}^2(r)=2\beta_{\phi} C_3 \frac{GM_*}{r_0}(\frac{r}{r_0})^{-1},~~~~~~~~~~~~~~~
\end{equation}

\begin{equation}\label{eq:cz}
c_{z}^2(r)=2\beta_{z} C_3 \frac{GM_*}{r_0}(\frac{r}{r_0})^{-1},~~~~~~~~~~~~~~~~~~~
\end{equation}

\begin{equation}\label{eq:h}
H(r)=C_4 r_0(\frac{r}{r_0}),~~~~~~~~~~~~~~~~~~~~~~~~~~~~~~~~~~~
\end{equation}
where $\Sigma_0$ and $r_0$ are denoted to provide the non-dimensional form of equations. By substituting the self-similar solutions (\ref{eq:s})$-$(\ref{eq:h}) into the basic equations (\ref{eq:continuity})$-$(\ref{eq:energy}) and (\ref{eq:hydrostatic}), we have
\begin{equation}\label{eq:s}
-C_0 C_1+\dot{m}=0,~~~~~~~~~~~~~~~~~~~~~~~~~~~~~~~
\end{equation}

\begin{equation}
-\frac{1}{2}{C_1}^2={C_2}^2-1-(s-\frac{3}{2})(1+\beta_{\phi}+\beta_{z})C_3-2\beta_{\phi}C_3,
\end{equation}

\begin{equation}
C_1=3\alpha(s+\frac{1}{2})\sqrt{C_3}C_4+\frac{2s\ell^2\dot{m}}{C_0},~~~~~~~~~~~~~~~~~
\end{equation}

\begin{eqnarray}\label{eq:e}
\big(\frac{1}{\gamma-1}+s-\frac{3}{2}\big)C_1 C_3=f\alpha\sqrt{C_3}C_4\Bigg(\frac{9}{4}{C_2}^2~~~~~~~~~
\nonumber\\+\frac{1}{2P_m}\Big( \beta_{\phi}(s-\frac{1}{2})^2+\beta_z(s-\frac{5}{2})^2 \Big)C_3\Bigg)-\frac{s\zeta \dot{m}}{4C_0},
\end{eqnarray}

\begin{equation}
{C_4}^2=C_3(1+\beta_{\phi}),~~~~~~~~~~~~~~~~~~~~~~~~~~~~~
\end{equation}
where the non-dimensional mass accretion rate $\dot{m}$ is defined as $\dot{M}_0/(2\pi r_0 \Sigma_0 \sqrt{GM_*/r_0})$. After some algebraic manipulations, a fourth-order algebraic equation is obtained for $C_4$:
\begin{eqnarray}\label{eq:mainequation}
\frac{81f\alpha^2(s+\frac{1}{2})^2}{8(1+\beta_{\phi})(1-2s\ell^2)^2}{C_4}^4+\Bigg[\frac{3(s+\frac{1}{2})}{(1+\beta_{\phi})(1-2s\ell^2)}~~~~~~~
\nonumber\\\big(\frac{1}{\gamma-1}+s-\frac{3}{2}\big)-\frac{9f}{4(1+\beta_{\phi})}\Big( (s+\frac{3}{2})(1~~~~~~~~~~~
\nonumber\\~~~~~~~+\beta_{\phi}+\beta_{z})+2\beta_{\phi} \Big)-\frac{f}{2P_m(1+\beta_{\phi})} \Big( \beta_{\phi}(s-\frac{1}{2})^2
\nonumber\\~~~~+\beta_{z}(s-\frac{5}{2})^2 \Big) \Bigg]{C_4}^2 -\frac{9}{4}f+\frac{3s\zeta(s+\frac{1}{2})}{4(1-2s\ell^2)}=0.
\end{eqnarray}
We are able to determine easily other flow quantities as a function of $C_4$. Their dependence on $C_4$ is
\begin{equation}\label{eq:sigma}
C_0=\frac{\sqrt{1+\beta_{\phi}}(1-2s\ell^2)}{3\alpha(s+\frac{1}{2})}\dot{m}{C_4}^{-2},~~~~~~~~~~~~~~~~~~~~~~~
\end{equation}

\begin{equation}\label{eq:radialvelosity}
C_1=\frac{3\alpha(s+\frac{1}{2})}{(1-2s\ell^2)\sqrt{1+\beta_{\phi}}}{C_4}^2,~~~~~~~~~~~~~~~~~~~~
\end{equation}

\begin{eqnarray}\label{eq:angularvelosity}
C_2=\Bigg\{ 1+\frac{1}{1+\beta_{\phi}}\Big[ (s-\frac{3}{2})(1+\beta_{\phi}+\beta_{z})+2\beta_{\phi} \Big]{C_4}^2
\nonumber\\-\frac{9\alpha^2(s+\frac{1}{2})^2}{2(1+\beta_{\phi})(1-2s\ell^2)^2}{C_4}^4 \Bigg\}^{1/2},
\end{eqnarray}

\begin{equation}\label{eq:pressure}
C_3=\frac{1}{1+\beta_{\phi}}{C_4}^2.~~~~~~~~~~~~~~~~~~~~~~~
\end{equation}
When the large-scale magnetic field has only the toroidal component, equation (\ref{eq:mainequation}) and the similarity solutions (\ref{eq:sigma})-(\ref{eq:pressure}) reduce to results of \cite{Faghei2012MNRAS}. In the absence of magnetic diffusivity, one can obtain the findings of \cite{Mosallanezhad2013} without the radial field component. By setting $\beta_{\phi}=\beta_z=0$, our solutions tend to the solutions presented by \cite{Shadmehri2008} when thermal conduction is ignored.

\section{Results}
In order to examine the effects of outflows and large-scale magnetic field on the properties of a resistive ADAF, we solve the equation (\ref{eq:mainequation}) numerically. To investigate the disc properties physically, only real roots corresponding to positive ${C_2}^2$ must be adopted. We illustrate the disc variables as a function of inverse magnetic Prandtl number, i.e. ${P_m}^{-1}$. In this paper, the magnetic Prandtl number is assumed to be in the range from 0.1 to 10 \citep{Lesur2007}. Note that the magnetorotational instability in this range is exited \citep{Kapyla2011}. We also set $\dot{m}=0.1$, $f =\ell=\zeta=1$, $\gamma=4/3$, $\alpha =0.1$, $\beta_\phi=\beta_z=1.0$, and $s=0.2$ unless otherwise is stated.

\begin{figure*}
\includegraphics[scale=1.0]{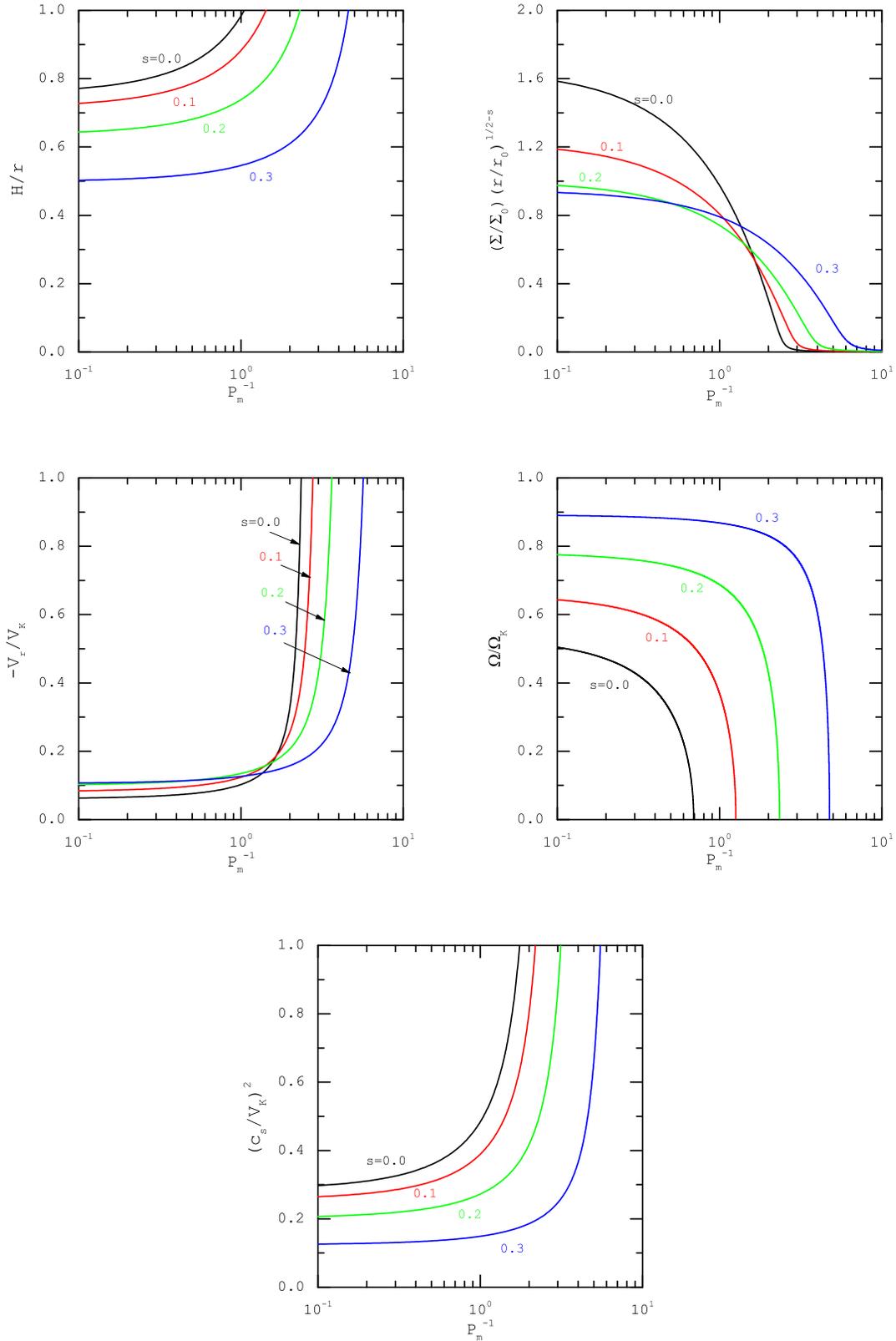}
\caption{Profiles of the physical variables of the accretion disc versus for different values of $s$, as labeled. It is assumed that and $\alpha=0.1$, $\ \beta_{\phi}=\beta_z=1.0$, $\zeta=\ell=f=1.0$.}\label{fig:f1}
\end{figure*}

\begin{figure*}
\includegraphics[scale=1.0]{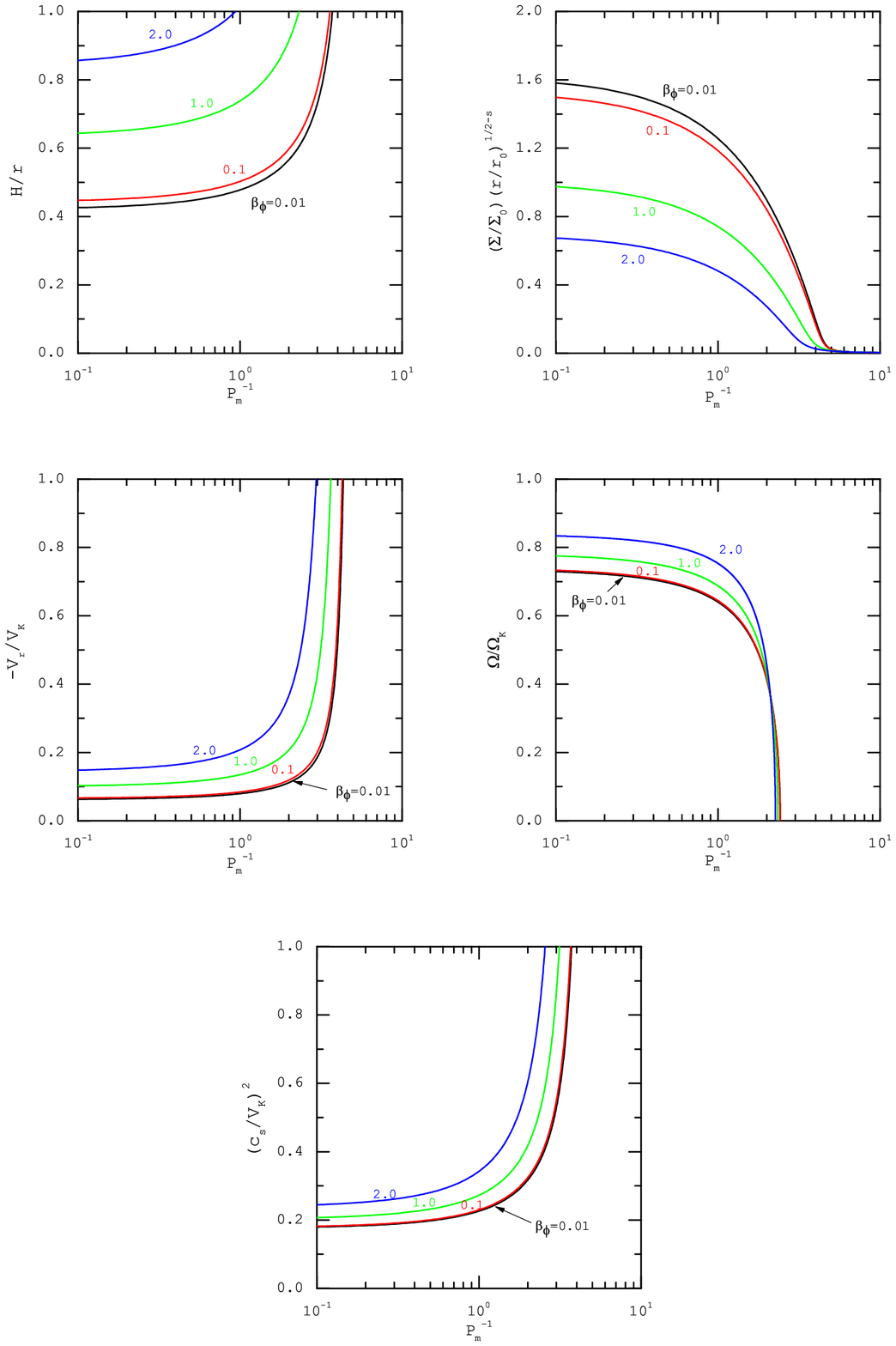}
\caption{Similar to Fig. \ref{fig:f1}, but for $s=0.2$ and different values of $\beta_{\phi}$ (as labeled).}\label{fig:f2}
\end{figure*}

\begin{figure*}
\includegraphics[scale=1.0]{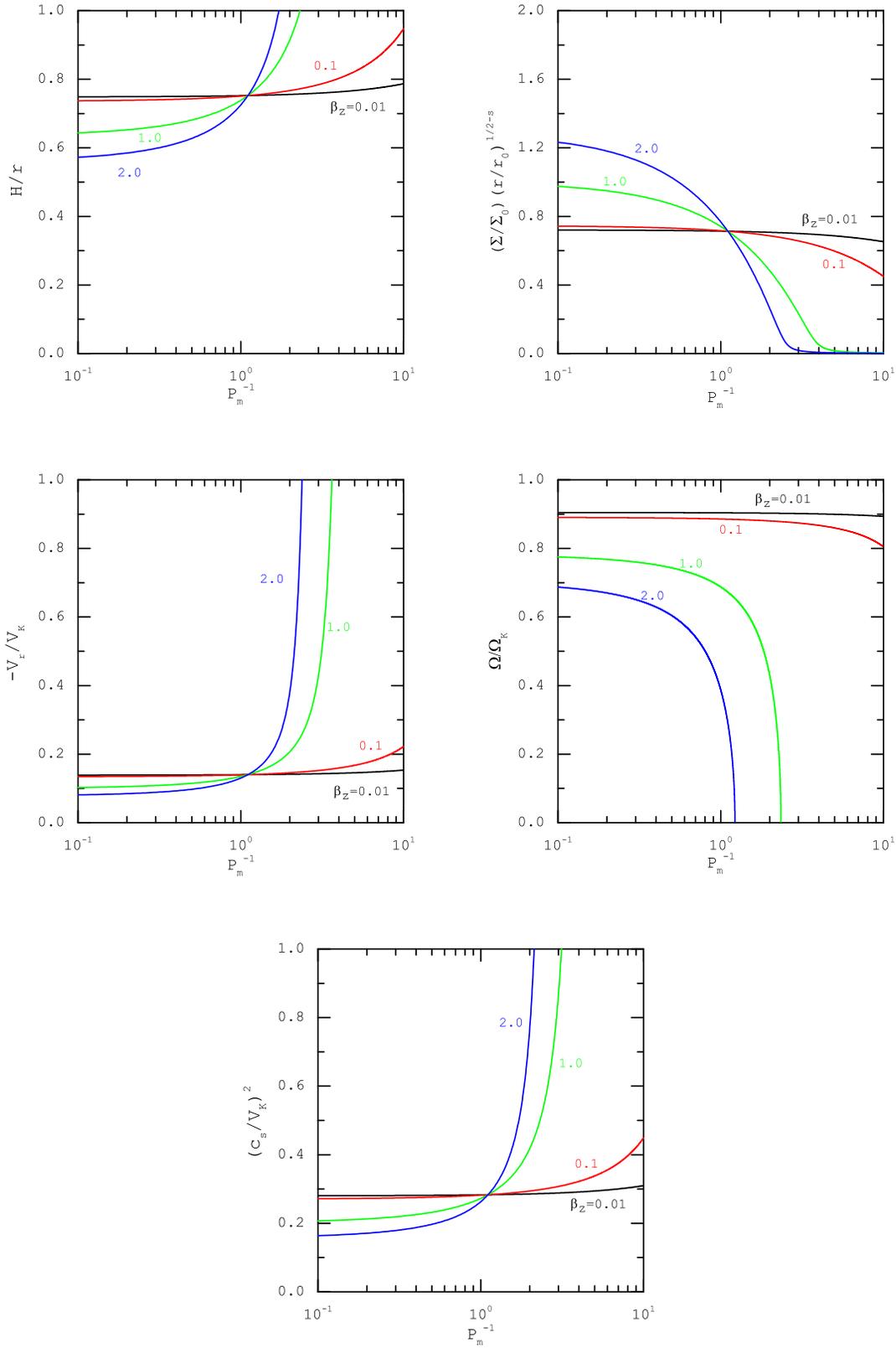}
\caption{Similar to Fig. \ref{fig:f1}, but for $s=0.2$ and different values of $\beta_z$ (as labeled).}\label{fig:f3}
\end{figure*}

In Fig. \ref{fig:f1}, the behavior of physical variables versus ${P_m}^{-1}$ is illustrated for different values of the power law index $s$. According to the momentum conservation, the term $1-2s\ell^2$ must be greater than zero (see also equation (\ref{eq:sigma})). Hence, the value of $s$ for a disc with $\ell=1$ lies between 0 and 1/2. Since the power law index $s$ for a case with $\alpha=0.1$ cannot be greater than 0.4 \citep{Yang2014}, we assume that the value of $s$ is in the range of 0.0 (a case without outflow) to 0.3 (a case with moderate outflow). In top left panel, the dimensionless thickness of disc $H/r$ is illustrated. In the present paper, the outflows are assumed to extract the energy from the disc. The removal of energy by outflows reduces the disc temperature. According to equation (\ref{eq:pressure}), the presence of outflows causes the disc thickness to decrease as revealed through previous numerical calculations \citep[e.g.][]{Faghei2012MNRAS,Ghoreyshi2020}. The ratio of $H/r$ rises with decreasing the magnetic Prandtl number. This means that an increase in the magnetic diffusivity, and therefore in the resistive heating, can lead to a higher gas temperature. Consequently, the enhancement of the gas temperature can puff up the accretion disc and the disc gets thicker. Noting that the $C_4-$dependence of density in equation (\ref{eq:sigma}), we also find that the surface density of the disc reaches a maximum value at the highest adopted value of $P_m$. In the high-$P_m$ limit, the surface density of a disc without outflow is more than that in a disc with moderate wind. Indeed, the outflows by extracting the disc material can lead to a reduction in the surface density. When the limit of small-$P_m$ is considered, however, the surface density of the disc in the absence of outflows becomes less. At $P_m=0.5$, for example, the surface density for a case $s=0.3$ is about twice the value in $s=0.0$. But \cite{Faghei2012MNRAS} showed that the net effect of the outflow on the surface density is independent of the magnetic Prandtl number. We find that such a difference in results is due to the presence of vertical component of magnetic field. The reason is as follows. According to equation (\ref{eq:e}), the radial velocity depends sensitively on the vertical magnetic field and the magnetic Prandtl number (see also the profile of the infall velocity in Fig. \ref{fig:f3}). The dependence of the surface density on the infall velocity (equation (\ref{eq:s})) leads to such a trend for the density profile.

As mentioned earlier, the presence of magnetic resistivity can lead to a thicker disc. According to equation (\ref{eq:radialvelosity}), the infall velocity is expected to increase due to an enhancement of the magnetic resistivity. Our results indicate that the radial velocity rises significantly in about $P_m=0.5$. At the higher magnetic Prandtl number, however, the infall velocity of the disc material remains almost unchanged. Although the outflows in this limit cannot significantly affect the radial velocity, the role of outflows in the modification of angular velocity is very important \citep[see also][]{Mosallanezhad2013}. The reason is as follows. In this paper, the outflows are assumed to extract the angular momentum. The angular momentum extracted from the disc enhances the angular velocity. We find that the rotational velocity depends not only on the the wind strength, but also on the magnetic Prandtl number. A decrease in the magnetic prandtl number causes the disc to rotate with a slower rate. Such a reduction in the rotational velocity could result from the increase of disc thickness with the magnetic resistivity (see equation (\ref{eq:angularvelosity})). One can see that there is a specific magnetic prandtl number in which the disc reaches a non-rotating limit. The value of such a specific magnetic prandtl number depends on the power law index $s$ because the rotational velocity is highly sensitive to the wind strength and the disc thickness. One find when stronger winds exist, the disc tends to a non-rotating limit in smaller $P_m$. In bottom panel of Fig. \ref{fig:f1}, the sound speed is illustrated. As we mentioned, this speed and therefore the disc temperature depend on the parameters $s$ and $P_m$. The sound speed can approach the Keplerian speed when ${P_m}^{-1}$ exceeds unity. The reduction in disc temperature due to an increase in the value of $s$ and $P_m$ is in good agreement with findings of \cite{Faghei2012MNRAS}.

The influence of $\beta_{\phi}$ is displayed in Fig. \ref{fig:f2}. In this figure, the power law index $s$ is 0.2 and $\beta_{\phi}$ can vary from 0.01 (i.e., ${c_{\phi}}^2=0.02{c_s}^2$) to 2.0 (i.e., ${c_{\phi}}^2=4{c_s}^2$), as labeled. Other input parameters are similar to Fig \ref{fig:f1}. In Fig. \ref{fig:f2}, higher values of $\beta_{\phi}$ are not considered because the ratio $H/r$ exceeds unity (see top left panel of Fig. \ref{fig:f2}). The dependence of the magnetic heating on the azimuthal field component implies that the stronger fields enhance the magnetic heating and the gas temperature. With increasing the temperature (or the sound speed) due to the toroidal component of the magnetic field, therefore, the disc becomes thicker. These results are in good agreement with previous works \citep[see][]{Ghoreyshi2020}. We find that at $P_m=10$, for example, the enhancement of $\beta_{\phi}$ from 0.1 to 2.0 causes the ratio $H/r$ to rise by two orders of magnitude. When the vertical field component is ignored, our results would be similar to findings of \cite{Faghei2012MNRAS}. Our results indicate that a stronger toroidal component can enhance the infall and the rotational velocities. A combination of equations (\ref{eq:radialvelosity}) and (\ref{eq:pressure}) exhibits that the radial velocity is proportional to $\sqrt{1+\beta_{\phi}}C_3$. Thus, the material can move radially inward with faster rate when a stronger azimuthal field component is considered. In order to explore the role of toroidal field component in the rotational velocity, it is helpful to compare the second term with the third one in equation (\ref{eq:angularvelosity}). For our input parameters, the value of the second term, i.e. the gradient of the total pressure plus the magnetic stress force, rises with increasing $\beta_{\phi}$. But the third term reduces as $\beta_{\phi}$ increases. Since the value of third term is smaller than the value of second one, the rotational velocity rises for stronger magnetic field. The previous studies of a resistive ADAF also showed that the toroidal component of the magnetic field can increase these velocities \citep{Faghei2012MNRAS}. Note that the specific magnetic prandtl number in which the disc has a non-rotating motion is almost independent of the input parameter $\beta_{\phi}$. A comparison between Figures \ref{fig:f1} and \ref{fig:f2} shows that the role of wind in the changes of velocities is more important.
One can expect that an increase in $\beta_{\phi}$ leads to a reduction in the surface density which can be because of the increase of radial velocity (see equation (\ref{eq:s})). Even if the radial field component is considered, the surface density decrease with increasing $\beta_{\phi}$ \citep[see][]{Mosallanezhad2013}.

Fig. \ref{fig:f3} displays the obtained results of investigation of the role of $\beta_z$ when $s=0.2$. Here, parameter $\beta_z$ lies between 0.01 and 2.0, as labeled, and other parameters are similar to Fig. \ref{fig:f1}. Under these condition, we find that the effect of vertical field component on almost all variables depends on the magnetic prandtl number. It results from the dependence of magnetic heating on the vertical field component. A comparison between the two terms of the magnetic heating exhibits that the coefficient related to the vertical field component is greater then that related to the toroidal component of the magnetic field. Thus, the influence of vertical field component can be more significant. In Fig. \ref{fig:f3}, one can see that the behavior of disc variables for different values of $\beta_z$ changes at a specific magnetic prandtl number about ${P_m}^{-1}=1.1$. When ${P_m}^{-1}<1.1$, for example, the ratio $H/r$ drops as the input parameter $\beta_z$ rises. But this trend would be reversed if ${P_m}^{-1}>1.1$. Since the rotational velocity is highly sensitive to the vertical field component (equation (\ref{eq:angularvelosity})), this velocity among the variables examined in this paper does not obey this trend. For our input parameters, the sign of the second term in equation (\ref{eq:angularvelosity}) is negative and an increase in $\beta_z$ results in a greater absolute value for this term. Hence, a larger $\beta_z$, unlike $\beta_{\phi}$, causes the disc material to rotate with a slower rate \citep[see also][]{Cao2013,Cao2014}. \cite{Mosallanezhad2013} also showed that there is a such trend for the rotational velocity in the absence of ohmic dissipation. For adopted range of $P_m$, the disc reaches a non-rotating limit when the vertical component of the magnetic field is strong enough. We also examine the effect of wind strength on $\Omega(P_m)$. Our results indicate that the power law index $s$ plays no role in vanishing the rotational velocity if the vertical component of the field is absent. This means that $\Omega(P_m)$ only due to the presence of strong vertical field falls to zero (compare all plots of $\Omega(P_m)$ in our model). Although the outflow is not the agent of vanishing $\Omega(P_m)$, its strength can affect the magnitude of the magnetic prandtl number $P_m$ in which the non-rotating limit occurs.

\begin{figure*}
\includegraphics[scale=1.0]{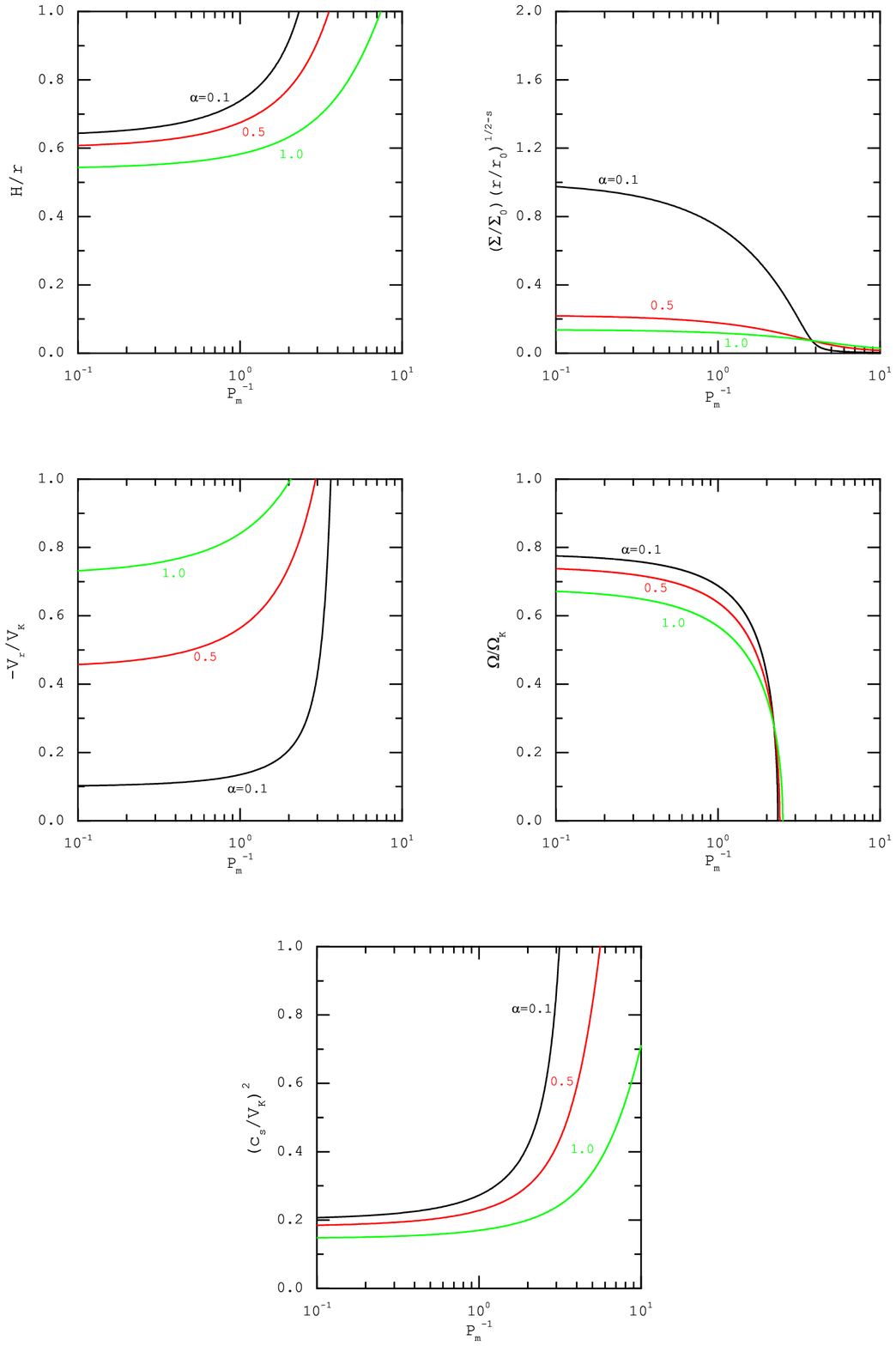}
\caption{Similar to Fig. \ref{fig:f1}, but for $s=0.2$ and different values of $\alpha$ (as labeled).}\label{fig:f4}
\end{figure*}

As the next step toward a more complete description of resistive discs, we explore the role $\alpha$ in the disc properties (see Fig. \ref{fig:f4}). According to equation (\ref{eq:eta}), we expect that the parameter $\alpha$ affects significantly the magnetic prandtl number and therefore the disc variables. In Fig. \ref{fig:f4}, the parameter $\alpha$ is assumed to vary from 0.1 to 1.0. We studied the influence of smaller values of $\alpha$, e.g. 0.01, and found that the difference between $\alpha=0.01$ and 0.1 is not important \citep[see also][]{Beckert2000}. Except the radial velocity, increasing the viscosity parameter $\alpha$ can reduce the variables examined in this work. We obtain the ratio of total heating rate to cooling rate as a function of $P_m$ and of $\alpha$. The cooling rate dominates the total heating rate if this ratio is below unity. As seen in previous figures, this ratio is greater than 1 and increases as the magnetic resistivity becomes more important. Furthermore, this ratio falls as the viscosity parameter grows. Hence, the temperature and the thickness of disc decrease with increasing $\alpha$. Although the infall velocity is related to ${(H/r)}^2$, it is also proportional to the viscosity parameter (see equation (\ref{eq:radialvelosity})). Thus, an increase in the viscosity parameter enhances the radial velocity. Indeed, the discs with higher viscosity parameter are able to transport higher angular momentum, and therefore the disc material moves inward with faster rate. Subsequently, the increase of infall velocity leads to a lower surface density (see equation (\ref{eq:s})). For small values of the magnetic prandtl number, the decrease of surface density due to the viscosity parameter is not significant. But if the viscosity parameter increases from 0.1 to 1.0, the surface density falls by ten orders of magnitude at high-magnetic prandtl number limit. In such a limit of $P_m$, however, the change in $\alpha$ from 0.1 to 1.0 causes the infall velocity to increase by seven orders of magnitude. One can also see that the magnetic prandtl number in which the non-rotating limit occurs is nearly independent of the viscosity parameter.

\section{Summary and Discussion}

The existence of outflows and large-scale magnetic fields in the inner regions of an ADAF have been confirmed by numerical simulation. In such regions of ADAFs, the magnetic diffusivity can play an important role in determining the properties of ADAFs. In the present paper, therefore, we investigated the dynamics of resistive ADAFs with coexistent outflows and large-scale magnetic fields. The large-scale magnetic fields were assumed to have $z$ and $\phi$ components. Here, the outflows can extract the mass, the angular momentum, and the energy from the disc. In the presence of outflows, numerical simulations and observations suggest a power-law function for the accretion rate whose power law index represents the wind strength. We assumed a steady state flow and presented the self-similar solutions that are described by a function of the radial distance.

Our self-similar solutions indicated that the rotational velocity of an ADAF threaded by large-scale magnetic field lines is always sub-Keplerian, in good agreement with previous works \citep[see][]{Ogilvie1998,Ogilvie2001,Cao2002,Cao2012}. However, the infall velocity in high-magnetic diffusivity limit can be close to the Keplerian speed. When the magnetic diffusivity becomes stronger, i.e. small-$P_m$ limit, the disc reaches a non-rotating limit and has a purely radial motion. Under such conditions, one can expect that the accretion rate of an ADAF rises. The changing accretion rate can alter the black hole luminosity \citep{Bu2019ApJ}. We also found that the rotational velocity depends strongly on the outflow and the vertical component of field. If the vertical field component is assumed to be weaker, the disc can rotate with
a faster rate. When outflow is present, the rotational velocity increases with increasing the outflow strength. As mentioned previously, the infall velocity in small-$P_m$ limit increases significantly due to the presence of magnetic diffusivity, so that the radial velocity of the disc material approaches the Keplerian speed. This increase is more significant when both vertical and azimuthal components of the large-scale magnetic field get stronger. Since the discs with higher viscosity parameter are able to transport higher angular momentum, the disc material moves inward with faster rate. This can lead to a high mass accretion rate.

We found that the surface density generally tends to zero in the high-magnetic diffusivity limit, unless the vertical field component is weak. Since the infall velocity, and therefore accretion rate increase at this limit, the surface density of a resistivity disc decreases significantly. Any change in the density profile can affect the emitted spectrum of an accretion flow. Although the surface density decreases with the increase of azimuthal field component, the effect of vertical field component on the surface density depends on the magnetic diffusivity. In large-magnetic diffusivity limit, the role of vertical field component in the modification of density profile is similar to what we found for the azimuthal field component. In the opposite limit, however, the trend of vertical field component will be different.

Our results showed that the disc becomes thicker with increasing the magnetic diffusivity. The disc thickness decreases as the wind strength or the viscosity parameter increases, while the disc becomes thicker for stronger azimuthal field component. We also displayed that an increase in the wind strength (or the viscosity parameter) reduces the disc temperature, while stronger azimuthal field component enhances the temperature. The temperature of an ADAF decreases due to stronger vertical field component only if small-magnetic diffusivity limit is considered. Indeed, the role of vertical field component in decreasing the disc temperature depends on the magnetic diffusivity. A comparison between our results and pervious works shows that the effective agent of temperature reduction is outflows and the decreases of temperature due to the vertical field component is seen only in ideal MHD and/or non-ideal MHD with low resistivity.

Although the self-similar solutions presented in this work may be very simplified, these solutions help us to understand the physics of ADAFs around a black hole and the origin of some observational phenomena. One example is the strong variations in the infrared and X-ray bands of Sgr A*. These fluctuations referred to as flares occur almost every day \citep[e.g.][]{Hornstein2007,Dodds2009}. The origin of the flares may be the electrons that are heated to higher temperatures and accelerated into a relativistic power-law distribution. The heating and the acceleration of the electrons in hot accretion flows could result from the magnetic reconnection \citep{Dodds2009,Ding2010}. \cite{Takahashi2013} showed that the enhancement of the resistivity causes the magnetic field lines to start to reconnect. Hence, the presence of the resistivity can lead to the heating and the acceleration of electrons. Furthermore, \cite{Ding2010} suggested that the resistive heating may be helpful to heat the electrons to higher temperatures, as we found in this work.

\begin{acknowledgements}
The author would like to thank an unknown referee for a constructive report. Thanks to Dr. Mohsen Shadmehri for a helpful comment.
\end{acknowledgements}

\bibliographystyle{pasa-mnras}
\bibliography{reference}

\end{document}